# Intermediate Mass Higgs Study at $\gamma\gamma$ Colliders[*]

Isamu WATANABE

*Department of Physics, Ochanomizu University*

1-1 Otsuka 2-chome, Bunkyo-ku, Tokyo 112, Japan

### ABSTRACT

We present the efficient technique to extract the signal of the intermediate mass Higgs boson from the backgrounds at future $\gamma\gamma$ colliders. For a clear Higgs detection, it is important to fit the original electron accelerator energy depending on the Higgs mass, to set the polarization of the photon beams and to apply the efficient $b$ quark tagging method. We demonstrate the extraction of information of Higgs parameters and the new physics from the observable physical quantities. It is clearly shown that a future $\gamma\gamma$ collider will have a rich potential for study on the new physics, as well as the Higgs physics.

---

[*] Talk presented at INS Workshop "Physics of $e^+e^-$, $e^-\gamma$ and $\gamma\gamma$ collisions at linear accelerators"

# 1   Introduction

A great idea to convert a future $e^+e^-$ collider into a $\gamma\gamma$ collider [1] is improving its reality due to the development of the high energy accelerator technology. Great physical potentials of $\gamma\gamma$ colliders have been suggested in the recent ten years [2, 3].

The most important target of a $\gamma\gamma$ collider will be on the Higgs production [4]. This process via loops of charged massive particles has a large cross section at the Higgs mass pole, in general, and a $\gamma\gamma$ collider is expected to play a role of the "Higgs factory."

Depending on the Higgs mass $m_H$, the main decay mode of the Higgs particle changes, and thus, the detection technique should be chosen according to the Higgs mass. The standard model Higgs mainly decays into a $b\bar{b}$ pair if $m_H \lesssim 140$ GeV (light and intermediate mass Higgs), while it decays into mainly a $W^+W^-$ pair for heavier Higgs mass. The decay modes into a $ZZ$ pair is secondary for $m_H \gtrsim 160$ GeV. The light Higgs or the intermediate mass Higgs should be detected by tagging a $b$ quark pair, and the heavy Higgs should be identified by double $Z$-tagging [5].

In the original naïve argument on the light and the intermediate mass Higgs detection, what people expected was as follows: There is a continuum contribution coming from the tree-level diagram in the $b\bar{b}$ process, and thus it may be a background to the resonant Higgs production. However, the Higgs peak is so sharp as several keV in the standard model, one can reduce the continuum background by reconstructing the $b\bar{b}$ invariant mass. Furthermore, Higgs particle only can be produced when the total angular momentum of the colliding two photon $J$ is vanishing, while the continuum $b\bar{b}$ production is negligible when $J = 0$ in the massless limit of $b$ quark due to the helicity conservation. Thus extracting the Higgs signal in $\gamma\gamma \to b\bar{b}$ process was expected to be an easy job, since the polarizations of the colliding photon beams will be controlled well.

Recently, some objections to the above rough argument were discussed by several authors. Éboli *et al.* [6] pointed out that the 'resolved' processes can be contribute to the backgrounds because of the gluonic content in the photon [7]. They estimated the backgrounds from the resolved processes assuming a few sets of the partonic distribution functions which have been proposed, and found that the backgrounds are too large for the $\gamma\gamma$ collisions with the c.m. energy of the original $e^+e^-$ collider 500 GeV. Baillargeon [8] also presented the similar results for the original $e^+e^-$ energies 350 and 500 GeV. In these analyses, however, the adopted partonic distributions are not the ones in the polarized photon, since they have not been determined yet at present. Abraham [9] suggested that the energy ratio of the parton in the photon for the background processes which can compete with the Higgs signal is bounded in the narrow region, and he assumed that the polarization effect to the distribution function can be expressed by a simple factor depends on the polarization of the photon. Then he showed the background cross section may have the ambiguity of factor 2 as the polarization effect.

On the other hand, Borden *et al.* [10] pointed out that the processes $\gamma\gamma \to b\bar{b}g$ and $\to c\bar{c}g$ do not governed by the helicity suppression, and thus, these processes have large cross sections even in the high energy limit with $J = 0$. The above processes can be the backgrounds to the Higgs signal when one of the final parton is soft or collinear to the initial photon beam axis, or when two of the final partons are collinear each other. In their conclusion, the backgrounds from this process are large at $\sqrt{s_{e^+e^-}} = 200$ GeV, however, one can extract Higgs signal by applying several kinematical cuts. Similarly, Jikia *et al.* [11] computed the radiative corrections to the processes $\gamma\gamma \to b\bar{b}$ and $\to c\bar{c}$, and found that the backgrounds dominate over the Higgs signal at a $\sqrt{s_{e^+e^-}} = 250$ GeV machine.

There is another background source of $\gamma\gamma \to e^+e^-Z \to e^+e^-b\bar{b}$ [12]. It may be serious if the Higgs mass is near at the Z boson mass.

These recent arguments suggested that the original naïve guess on the Higgs detection should be reexamined. Unfortunately, these arguments did not take into account the realistic distributions of the luminosity and the polarizations at the $\gamma\gamma$ collider[1] . It has been pointed out that the polarization distribution hardly depends on the colliding c.m. energy ratio [13]. Since the Higgs signal cross section also depends on the polarizations of the colliding photons, one should perform the complete analysis with the realistic distributions of the luminosity and the polarizations. Borden *et al.* [14] performed an analysis on that manner, however, they assumed 250 GeV of the original $e^+e^-$ energy.

---

[1] Only the luminosity distribution is taken into account in Ref. [6].

All of the arguments already presented is based on the original $e^+e^-$ colliders with $\sqrt{s_{e^+e^-}} = 200$—$500$ GeV. Such machines are too energetic to detect an intermediate mass Higgs clearly.

In the present paper, we show the importance of adjusting the accelerator energy depending on the Higgs mass to extract the signal of the intermediate mass Higgs boson from the backgrounds efficiently at a future $\gamma\gamma$ collider, and then we demonstrate the high physical feasibility of the $\gamma\gamma$ collider. The organization of the paper is as follows: In section 2, we illustrate the distributions of the luminosity and the polarizations at the $\gamma\gamma$ colliders. The cross sections of the Higgs signal process and the backgrounds are estimated in the section 3. The physical quantities extractable from the observed event rates are examined in the section 4. Section 5 gives the conclusions.

## 2  Luminosity and Polarization Distributions

Theoretical estimations to the luminosity and the polarization distributions have been evaluated by many authors [1, 2, 3, 13]. The most convenient and simple estimation is assuming that the laser photon is scattered by the beam electron only once, and the scattering angle of the energetic photon in the laboratory frame is negligible. Recently, Ogaki *et al.* [15] performed a more realistic simulation of the beam conversion mechanism including the multi-scatterings of the electrons and the photons, however, their result depends on the detailed beam design of the $\gamma\gamma$ collider. Thus we here adopt the simple formulae as in Ref. [13] for the general physics analyses.

It should be commented on the collision energy limit of the $\gamma\gamma$ colliders. Neglecting the multi-scatterings in the beam conversion, the maximum ratio of the c.m. energy of the $\gamma\gamma$ collision to that of the original $e^+e^-$ collider is described as $x/(1+x)$, where $x$ is the ratio squared of the c.m. energy of the Compton scattering of the beam conversion to the electron mass. It was believed that $x$ should be set less than $2\sqrt{2}+2$ to avoid the by-production of an $e^+e^-$ pair from the collision of the energetic photon and the laser photon [1, 3]. Within this restriction, the $\gamma\gamma$ collision energy is limited to $2\sqrt{2}-2 \sim 0.828$ times the original $e^+e^-$ collider energy. According to the recent realistic simulation [15] shows the event rate of the by-production is small, and it may be able to adopt a larger $x$ than $2\sqrt{2} + 2$. We here stick into the traditional constraint, and assume $x = 2\sqrt{2} + 2$, despite of the recent observation.

The luminosity distributions of $\gamma\gamma$ collisions in both $J = 0$ and $J = 2$ can be found in Fig. 1. The polarizations of the electron beams and the lasers are set up so that the luminosity distribution in $J = 0$ collisions highly peaks at large energy fraction $z$. Since only the $J = 0$ collisions are responsible for the Higgs production, the Higgs production is efficient if the Higgs mass is at around the 0.8 times the original $e^+e^-$ collider energy. At the same time, the $q\bar{q}$ backgrounds which mainly come from $J = 2$ collisions are suppressed at $z \sim 0.8$. For the lighter mass Higgs, the Higgs production rate decreases, and the backgrounds grow up rapidly.

## 3  Signal vs. Backgrounds

Now we evaluate the event rate of the Higgs signal and the backgrounds in detail.

### 3.1  Signal

We assume the standard model Higgs with the intermediate mass for instance. The Higgs signal should be identified by $b\bar{b}$ pair with the invariant mass same as the Higgs mass within the detector accuracy.

(i). $\gamma\gamma \to H \to b\bar{b}$

Since the Higgs decay width is as sharp as 10 keV, the peak cross section can be approximated by Breit–Wigner formula,

$$\sigma_{\gamma\gamma \to H \to b\bar{b}} = 16\pi \frac{\Gamma_{\gamma\gamma}\, \Gamma_{b\bar{b}}}{(s_{\gamma\gamma} - m_H^2)^2 + m_H^2 \Gamma_H^2}\,, \tag{1}$$

where $s_{\gamma\gamma}$ is the $\gamma\gamma$ collision energy squared, $\Gamma$'s are the total and the partial decay widths of the Higgs and $m_H$ is the Higgs mass. The observable is not the shape of this cross section curve, but the number of the event rate $N_{\gamma\gamma \to H \to b\bar{b}}$, i.e. the convolute integral of the cross section with the $\gamma\gamma$ luminosity distribution,

$$\begin{aligned} N_{\gamma\gamma \to H \to b\bar{b}} &= \int d\sqrt{s_{\gamma\gamma}} \frac{dL_{\gamma\gamma}^{\pm\pm}}{dz} \frac{1}{\sqrt{s_{\gamma\gamma}}} \sigma_{\gamma\gamma \to H \to b\bar{b}} R(\Delta) \\ &= \frac{1}{L_{\gamma\gamma}} \frac{dL_{\gamma\gamma}^{\pm\pm}}{dz}\bigg|_{\text{pole}} \frac{1}{\sqrt{s_{e^+e^-}}} \left[ 8\pi^2 \frac{\Gamma_{\gamma\gamma} B_{b\bar{b}}}{m_H^2} \right] R(\Delta) L_{\gamma\gamma}, \end{aligned} \quad (2)$$

where $L_{\gamma\gamma}$ is the $\gamma\gamma$ luminosity, $L_{\gamma\gamma}^{\pm\pm}$ the $J=0$ part of the $\gamma\gamma$ luminosity, $B_{b\bar{b}}$ the $b\bar{b}$ branching ratio of the Higgs decay. The factor $R(\Delta)$ describes the fraction of the selected signal events within $m_H \pm \Delta$, and is expressed by the error function of the Gaussian distribution. We simply assumed $\Delta = 5$ GeV which is roughly one sigma of the detector resolution[2], and thus $R = 3/4$. The suffix 'pole' means to evaluate the value at $s_{\gamma\gamma} = m_H^2$. For the comparison of the signal with backgrounds, it is convenient to define the 'effective' cross section $\sigma^{\text{eff}}$,

$$\begin{aligned} \sigma^{\text{eff}}_{\gamma \to H \to \gamma\gamma} &= N_{\gamma\gamma \to H \to b\bar{b}}/L_{\gamma\gamma} \\ &= \frac{1}{L_{\gamma\gamma}} \frac{dL_{\gamma\gamma}^{\pm\pm}}{dz}\bigg|_{\text{pole}} \frac{1}{\sqrt{s_{e^+e^-}}} \left[ 8\pi^2 \frac{\Gamma_{\gamma\gamma} B_{b\bar{b}}}{m_H^2} \right] R(\Delta). \end{aligned} \quad (3)$$

The total luminosity $L_{\gamma\gamma}$ is estimated to be about 10 fb$^{-1}$/yr, by fitting to the simulation result for $z > 0.6$ by Ogaki *et al.* [15].

## 3.2 Backgrounds

The possible background processes are as follows:

(ii). $\gamma\gamma \to b\bar{b}$

(iii). $\gamma\gamma \to c\bar{c}$

(iv). $\gamma\gamma \to b\bar{b}(g)$

(v). $\gamma\gamma \to c\bar{c}(g)$

where gluon in the parentheses means it is emitted invisibly due to the beam pipe holes. The events with the $b\bar{b}$ invariant mass similar to the Higgs mass can be mixed with

---

[2] The signal-to-background ratio can be improved if we take larger $\Delta$. However, we made this simple choice in this analysis. See Ref. [21] for the optimum choice of $\Delta$.

– 6 –

the Higgs signals. The charm quarks may be misidentified with the bottom quarks in the vertex detector, however, the jets from a light quark or a gluon can be eliminated efficiently.

The 'effective' cross sections of the background processes can also be estimated as,

$$\sigma_{\text{BG}}^{\text{eff}} = \int_{m_H-\Delta}^{m_H+\Delta} d\sqrt{s_{\gamma\gamma}}\, \frac{d\, L_{\gamma\gamma}}{d\, z}\, \frac{1}{\sqrt{s_{\gamma\gamma}}}\, \sigma_{\text{BG}}\ . \tag{4}$$

## 3.3 Numerical Results

The computations of the effective cross sections are performed in numerical way by some FORTRAN programs with a subroutine package for the helicity amplitude evaluation HELAS [16] and with a Monte Carlo integration codes BASES25 [17]. Some experimental cuts are introduced:

- Both jets from the quark and the anti-quark should clearly be visible in the detector: *i.e.* $|\cos\theta_q|,\ |\cos\theta_{\bar{q}}| < 0.7$.

- The jet from the gluon should be invisible in the detector: *i.e.* $|\cos\theta_g| > 0.9$.

- Jets should be clearly isolated: *i.e.* $m_{ij}^2/s_{e^+e^-} > 0.02$, where $m_{ij}$ is the invariant mass of the two jets from partons $i$ and $j$.

- The missing transverse momentum and aplanarity due to invisible gluon should be small: *i.e.* $\not{p}_T < 10$ GeV and $||\phi_q + \phi_{\bar{q}}| - \pi| < 0.02$, where $\phi$'s are the azimuthal angles of the quark and the anti-quark.

- Sum of the momenta of the quark and the anti-quark should be satisfy a relation that should be hold if no gluon is emitted: *i.e.* $|\eta_{q\bar{q}}| < \log((2\sqrt{2}-2)/z_{q\bar{q}})$, where $\eta_{q\bar{q}}$ and $z_{q\bar{q}}$ are rapidity and energy fraction of the $q\bar{q}$ system, respectively.

We assume the original $e^+e^-$ collider energy to be 150 GeV, and adopt the three typical Higgs masses 90, 105 and 120 GeV. The obtained effective cross sections are summarized in Table 1. Just as the naïve guess from the luminosity distributions, the signal is remarkable at $m_H = 120$ GeV, and backgrounds are huge for smaller Higgs mass. It is found that the $q\bar{q}(g)$ backgrounds can be negligibly small.

The $b$-tagging technique with a vertex detector was found efficient to reject the light quark jets and gluon jets. We estimated the tagging efficiencies of the vertex detector from the simulation of the JLC detector on the jets at the $Z$ pole [18]. For the selecting criteria that requires at least 3 charged tracks in the vertex detector and at least 2.5 times distant impact parameter compared to the resolution, $b\bar{b}$-tagging efficiency is estimated to be 40%, while the 2.0% of the $c\bar{c}$ pairs survive the same cuts, where we mean that both of the quark and the anti-quark should be satisfy the criteria. The efficiency for the light quark pair events and $\gamma\gamma \to b(\bar{b})g$, $c(\bar{c})g$ events are $6.3 \times 10^{-6}$, $1.6 \times 10^{-3}$ and $3.5 \times 10^{-4}$, respectively, and thus these events can be negligible.

As mentioned above, only the events that both heavy quarks decay hadronically should be collected to estimate the $\gamma\gamma$ collision energy. The hadronic decay branching ratio of a $b$-flavored hadron is 75%, on the other hand, that of a $c$-flavored meson is 82% [19].

Multiplying the above tagging efficiencies and the hadronic branching ratio squared, the effective cross section become the values summarized in Table 2. Note here that the $c\bar{c}$ background is suppressed in this time because of the efficient $b$-tagging. For the suitably matched $m_H = 120$ GeV, the signal-to-background (S/B) ratio is as large as 21, and the statistical significance of the signal is 66 standard deviations for a year run.

## 4 Discussions

In this section, we briefly discuss on the physical feasibilities of a $\gamma\gamma$ collider on the determinations of the nature of the Higgs particle, in the case of the appropriate energy set up, *i.e.*, $\sqrt{s_{e^+e^-}} = 150$ GeV for $m_H = 120$ GeV. Through out the section, we assumed $\Delta = 5$ GeV and $\int dt\, L_{\gamma\gamma} = 10$ fb$^{-1}$.

### 4.1 Cross Section

For one year run of the collider, the expected number of the events survived the selection criteria described above is 218, while the backgrounds mingled in is expected to be 10 events. The estimated signal is 208 events, and its statistical error is $\sqrt{228} \sim 15$. Thus, the statistical error ratio of the signal cross section is only 7%.

## 4.2 Mass Determination

The Higgs mass can be measured by the $b\bar{b}$ invariant mass. Each signal event has an error due to the detector resolution, and the standard deviation of the $b\bar{b}$ invariant mass measurement to one event is 4 GeV for $m_H = 120$ GeV. With the 280 signal events, the resultant mass resolution is 0.3 GeV. This means that a $\gamma\gamma$ collider has the similar performance on the Higgs mass determination with future $e^+e^-$ colliders.

## 4.3 Two-Photon Decay Width

The two-photon decay width of Higgs $\Gamma_{\gamma\gamma}$ is quite important to look beyond the standard model, since the contributions of the heavy charged particles in the loop diagrams do not decouple, if the masses of such heavy particles are generated only by the present Higgs in interest.

Assuming the standard model value of $B_{b\bar{b}}$, one can extract the $\Gamma_{\gamma\gamma}$ from the event number,

$$\Gamma_{\gamma\gamma} = \frac{N_{\gamma\gamma \to H \to b\bar{b}}}{\frac{8\pi^2}{m_H^2} \left. \frac{d L_{\gamma\gamma}^{\pm\pm}}{d\sqrt{s_{e^+e^-}}} \right|_{\text{pole}} R(\Delta)\, B_{b\bar{b}}} \,. \qquad (5)$$

The statistical error of the width is again 7%, neglecting the error from the luminosity distribution, and thus the error of the absolute value of the amplitude sum of the all contributing loop diagrams is 4%.

## 4.4 Sensitivity to New Physics

We now demonstrate the sensitivity to the new physics of a $\gamma\gamma$ collider. We assume an additional heavy generation of quarks and leptons, whose masses are generated by the Yukawa couplings with the present intermediate mass Higgs. For a simplicity, all masses of the quarks and charged leptons are assumed to be degenerated and heavy enough compared with the present experimental energy scale, since it results the least change from the 3-generation standard model.

The two-photon decay width can be evaluated as follows [20]:

$$\Gamma_{\gamma\gamma} = \frac{\alpha^3}{256\pi^2 \sin^2\theta_W} \frac{m_H^3}{m_W^2} \left|\sum_i N_i e_i^2 F_i\right|^2, \tag{6}$$

where $i$ is a suffix of the contributing charged and massive particles, $N_i$ the degree of freedom like the color factor, $e_i$ the charge of the particle $i$, and $F_i$ the function depends on the spin of $i$, as well as the mass ratio of the particle $i$ and Higgs (See Ref. [20] for details). For the 3-generation standard model, the sum of the amplitude $\sum_i N_i e_i^2 F_i$ is $6.44 - 0.09i$. Assuming the fourth generation, the amplitude reduces to $2.88 - 0.09i$. If the nature realizes the 3-generation standard model, the fourth generation model is rejected by 15 sigma of the statistical significance. On the other hand, if there is the fourth generation, the signal event number decreases to 42 according to the change of the amplitude, and thus the resolution to the amplitude grows to 9%. The measured absolute value of the amplitude will be at $2.88 \pm 0.27$. Even in this case, the 3-generation standard model is rejected by 13 sigma.

In the case with the fourth generation, the mass lower bound of the degenerated generation is, unfortunately, 112 GeV at one standard deviation, which should be a target of the direct search by $e^+e^-$ colliders with $\sqrt{s} \geq 224$ GeV. The sensitivity to the mass of the new physics is disappointedly poor, since the function $F_i$ saturates immediately at below the threshold of open $i$-particle pair production.

The poor sensitivity to the mass scale of the new physics allows to take the infinitesimal limit of the new particle mass in the amplitude evaluation, and thus the new physics contribution to the amplitude is simply approximated to be $\sum_i N_i e_i^2 F_i^\infty$. Here $F_i^\infty$ is a number depends on the spin of the new particle, i.e., $F_i^\infty = +7$ for a vector particle, $-4/3$ for a spinor particle and $-1/3$ for a scalar particle. Because of a good accuracy to the absolute value of the amplitude, one can expect that many types of the new physics models with different $\sum_i N_i e_i^2$ can be resolved experimentally. It is quite important that a future $\gamma\gamma$ collider can distinguish many distinct models with higher energy new particles.

For the case of the new heavy particle whose mass is generated only partially by the present Higgs particle, just like the scalar fermions in the supersymmetric models, the additional two-photon decay amplitude depends on the fraction of the Higgs contribution in the mass generation mechanism. In such a situation, there may be some sensitive regions on the mass scale of the new physics.

## 5    Conclusions

We have presented the efficient technique to extract the signal of the intermediate mass Higgs from the backgrounds at future $\gamma\gamma$ colliders. The important keys are; (1) to fit the original $e^+e^-$ machine energy to be $1/0.8$ times to the Higgs mass, (2) to set the polarizations of the photon beams such that the luminosity distribution peaks at around $z \sim 0.8$ in $J = 0$ $\gamma\gamma$ collisions, and (3) to apply the efficient $b$-tagging method by the vertex detector. We have also demonstrated the derivation of the information of the new physics, as well as the Higgs parameters, from the observable physical quantities. It has been clearly shown that a future $\gamma\gamma$ collider will have rich physical feasibilities for both studies on the Higgs and on the new physics.

More concrete analysis will be available in the near future [21].

## Acknowledgments

The author would like to acknowledge valuable discussions with Professor K. Hagiwara, Professor R. Najima, Dr. T. Kon, Dr. Y. Sugimoto, Dr. T. Takahashi, Dr. K. Tesima, Mr. T. Ogaki, Mr. Y. Yasui and Ms. F. Kanakubo. The author is grateful to Professor A. Sugamoto and the members of his laboratory in Department of Physics, Ochanomizu University, and Professor T. Muta for constant encouragements. A part of the numerical calculation was performed on a computer FACOM M780 at Institute for Nuclear Study, University of Tokyo.

# Tables

**Table 1** Effective cross sections of Higgs signal process and background processes. Experimental cuts described in the text are applied.

| process | Higgs mass | | |
|---|---|---|---|
| | 90 GeV ($z$=0.6) | 105 GeV (0.7) | 120 GeV (0.8) |
| $\gamma\gamma \to H \to b\bar{b}$ | 8.4 fb | 38 fb | 92 fb |
| $\gamma\gamma \to b\bar{b}$ | 31 fb | 18 fb | 2.9 fb |
| $\gamma\gamma \to c\bar{c}$ | 500 fb | 270 fb | 23 fb |
| $\gamma\gamma \to b\bar{b}(g)$ | 0.5 fb | 0.1 fb | 0.0 fb |
| $\gamma\gamma \to c\bar{c}(g)$ | 7.4 fb | 1.2 fb | 0.0 fb |

**Table 2** Effective cross sections multiplied by the detection efficiency and the hadronic branching ratio squared of the Higgs signal process and background processes. Same cuts as Table 1 are applied.

| process | Higgs mass | | |
|---|---|---|---|
| | 90 GeV ($z$=0.6) | 105 GeV (0.7) | 120 GeV (0.8) |
| $\gamma\gamma \to H \to b\bar{b}$ | 1.9 fb | 8.6 fb | 20.8 fb |
| $\gamma\gamma \to b\bar{b}$ | 7.1 fb | 4.1 fb | 0.7 fb |
| $\gamma\gamma \to c\bar{c}$ | 6.7 fb | 3.7 fb | 0.3 fb |
| $\gamma\gamma \to b\bar{b}(g)$ | 0.1 fb | 0.0 fb | 0.0 fb |
| $\gamma\gamma \to c\bar{c}(g)$ | 0.1 fb | 0.0 fb | 0.0 fb |
| sum of backgrounds | 14.0 fb | 7.8 fb | 1.0 fb |
| S/B ratio | 0.13 | 1.1 | 21 |
| $S/\sqrt{B} \times \sqrt{10\ \text{fb}^{-1}}$ | 1.6 | 9.7 | 66 |

# Figure

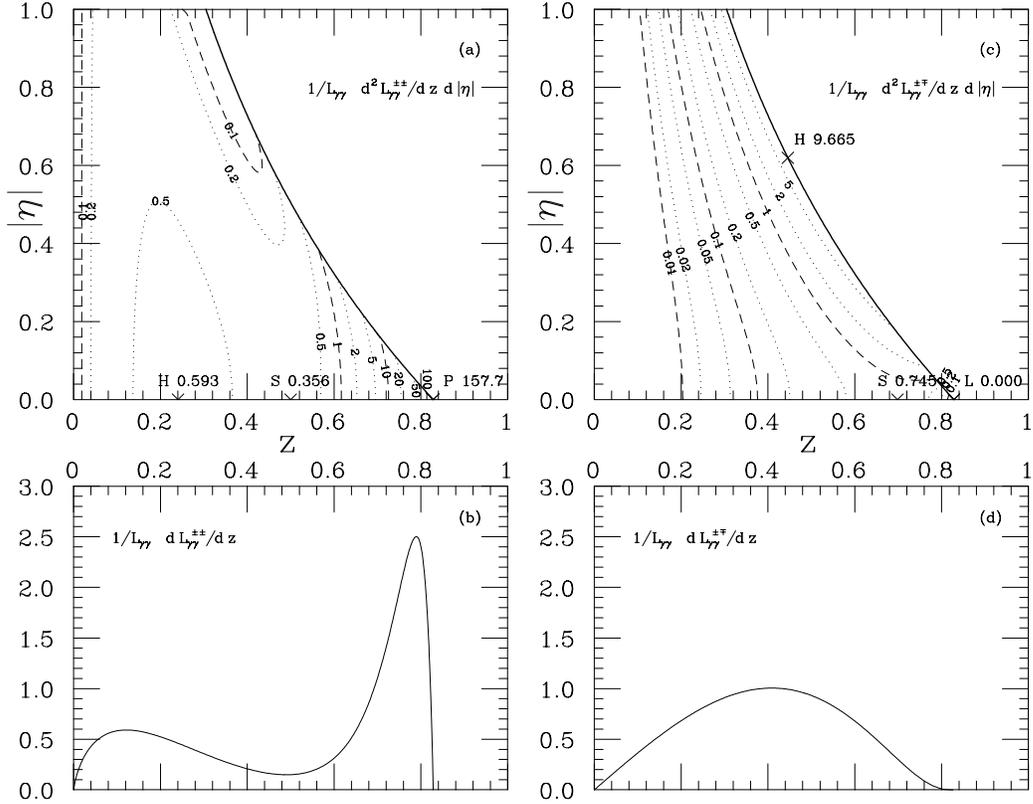

Figure 1: Luminosity distributions of a $\gamma\gamma$ collider. Left (right) graphs illustrate the $J = 0$ ($J = 2$) component. Upper two show the contours in the $z$-$\eta$ plane, while the bottom two are distributions integrated out on the rapidity $\eta$. The luminosities given here are normalized so that the total luminisity of the two spin combinations is unity.